\documentclass[pra,reprint,superscriptaddress,twocolumn,showkeys,amsmath,amssymb,a4paper]{revtex4-1}
\usepackage{graphicx}
\usepackage{subfigure}
\usepackage{dcolumn}
\usepackage{bm}
\usepackage{amsmath}
\usepackage{amssymb}
\usepackage{txfonts}
\usepackage{multirow}
\usepackage{subfloat}
\usepackage[colorlinks,linkcolor=blue,anchorcolor=blue,citecolor=blue]{hyperref}
\usepackage{blindtext}
\usepackage{booktabs}
\usepackage{threeparttable}
\usepackage{comment}
\usepackage{times}
\usepackage{amsmath,amstext}
\mathcode`<="4268 \mathcode`>="5269

\begin{document}

\preprint{AIP/123-QED}
%\begin{frontmatter}
%\begin{titlepage}
%\begin{center}
%\title{Self-consistent field method in the density-matrix functional theory with a correlation energy functional based on the maximum entropy corresponding to the Fermi-Dirac distribution}
\title{Reply to Comment on ``Self-Consistent-Field Method for Correlated Many-Electron Systems with an Entropic Cumulant Energy'', arXiv2202.05532v1}

\author{Jian Wang}
\affiliation{School of Science, Huzhou University, Zhejiang 313000, China}
\author{Evert Jan Baerends}
\affiliation{Afdeling Theoretische Chemie, FEW, Vrije Universiteit, De Boelelaan 1083, 1081 HV Amsterdam, The Netherlands}

%\date{\today}
%\pacs{31.25.Jf}{Electron correlation calculations for atoms and ions: excited states}
%\pacs{31.10.+z}{Theory of electronic structure, electronic transitions, and chemical binding}
%\pacs{31.15.Ar}{Ab initio calculations}
%\pacs{31.25.Jf,31.10.+z,31.15.Ar}
\begin{abstract}
Recently we proposed an information entropy based method for electronic structure calculations within the density-matrix functional theory(DMFT) (Phys. Rev. Lett. 128, 013001), dubbed as $i$-DMFT. 
Comments have been raised regarding the accuracy of the one-particle density-matrix compared to that from the wave function, and the universality of the functional. We address these questions and the problematic case of HeH$^+$. 
 \end{abstract}

\maketitle

Ding, Liebert and Schilling (DLS) submitted a critical comment \cite{SchillingComment} on our recent publication \cite{WangPRL2022} which describes a procedure called $i$-DMFT. 

DLS state that we make three claims regarding our method: (i) that it is a method \textit{within DMFT} (ii) that it is capable of describing accurately molecules at various geometries, in particular the dissociation limit; (iii) that it is based on a distinctive information-theoretic approximation of the 2RDM cumulant. They then argue that these claims are unjustified. We address their points in order.\\

(i) As for the first claim, from the further elaboration it is clear that the authors mean that we claim to offer a method that tries to reproduce the exact one-particle reduced density matrix (1RDM) using a universal functional of the natural orbitals (NOs) and occupation numbers (ONs). But we do not claim to approximate the exact 1RDM and we clearly do not offer a universal functional and do not claim to do so.\\

Our method is clearly an orbital method, or self-consistent field method, as the Hartree-Fock and Kohn-Sham methods are. The crucial point, however, is that we employ fractional occupation numbers (a density matrix) in order to invoke the extra flexibility that is offered by the involvement of the ``virtual'' orbitals. So it is a density matrix method, but evidently it does not try to reproduce the exact NOs and ONs. It is well-known that Gilbert's \cite{Gilbert:1975hw} one-electron equations for the NOs lead to degeneracy of all fractionally occupied  NOs, while we rely on the orbital energies of our method for providing the basis of the Fermi-Dirac occupation numbers. We also note that, while the (strongly) occupied NOs are close to occupied HF and KS orbitals, the first virtual NOs lack the typical characteristics that make traditional orbitals useful  (we call the weakly occupied spinorbitals, with occupations below 0.5, just virtual orbitals). The first virtual NOs, also in the case of weak (dynamic) correlation, are concentrated in the molecular bulk region. They actually may be more compact than the strongly occupied NOs, which is understandable from the role they play in electron correlation. This means they do not represent excited electrons. It is very hard to deduce the nature of excitations from the nature of the (usually many) NOs involved in expressing an excited state. One has to make a specific transformation to a different orbital basis (e.g.\ the so-called Natural Excitation Orbitals \cite{vanMeerGritsenkoBaerends2015}) in order to obtain an interpretation of an excitation in the familiar terms (e.g.\ $\pi \to \pi*$, $n \to \pi*$, Rydberg, metal-to-ligand, etc.). The Kohn-Sham model offers such a basis of very meaningful virtual orbitals (exact KS much better than LDA/GGA/hybrid functionals) that afford a description of many excitations as just orbital-to-orbital transitions \cite{vanMeerGritsenkoBaerends2014b}. With our orbital method we aim at such chemical/physical utility, in addition to computational expediency, not at reproducing NOs. Our method is therefore more akin to the efforts in density functional theory (DFT) to exploit the additional flexibility offered by the use of virtual orbitals, usually by some Fermi-Dirac type occupation scheme, cf.\ Refs \cite{BaerendsPRL2001,Gruning2003,ChaiJengDa2012,Baldsiefen2013,Grimme2015}. The advantage now is that the Fermi-Dirac occupation scheme is not heuristic but is determined variationally. 

Since we do not target the exact 1RDM, we do not have to (and do not try to) obey the generalized Pauli constraints \cite{Borland:1972bp,Klyachko2006,Altunbulak:2008kg}. A 1DM functional built upon Pauli constraints $1 \ge n_i \ge 0$ alone usually violates the generalized Pauli constraints for the 1RDM \cite{Lathiotakis2015}. \\
\\
(ii) We have demonstrated that remarkably accurate dissociation curves could be obtained for diatomic molecules with our Ansatz for the cumulant energy, $E_{\rm cum}=-\kappa S(\mathbf{n})-b$, giving for the electron-electron interaction energy
\begin{equation}\label{eq:Eee}
E_{ee}=Y-\kappa S(\mathbf{n})-b
\end{equation}
with $Y$ the direct and exchange parts of the energy
\begin{equation}\label{eq:Y}
Y=\frac{1}{2} \int \int \frac{\gamma(1,1)\gamma(2,2)-\gamma(1,2)\gamma(2,1)}{r_{12}} d1d2'.
\end{equation}
The fractional occupation numbers help us to solve the notorious molecular dissociation problem.  DFT has the problem that the electron density is rather insensitive to electron correlation. At near dissociation the local density on a weakly bonded fragment in practice is not revealing its entanglement with the rest of the system. Involvement of virtual orbitals solves this entanglement problem \cite{Baerends2020Janak}, which is also a motivation for the application of density-matrix functional theory (DMFT) in general. We have referred to related attempts in DFT \cite{BaerendsPRL2001,Gruning2003,ChaiJengDa2012,Baldsiefen2013,Grimme2015} to invoke virtual orbitals by fractional occupation schemes.\\

\begin{figure}[h]
 \begin{center}
  \includegraphics[width=.5\textwidth]{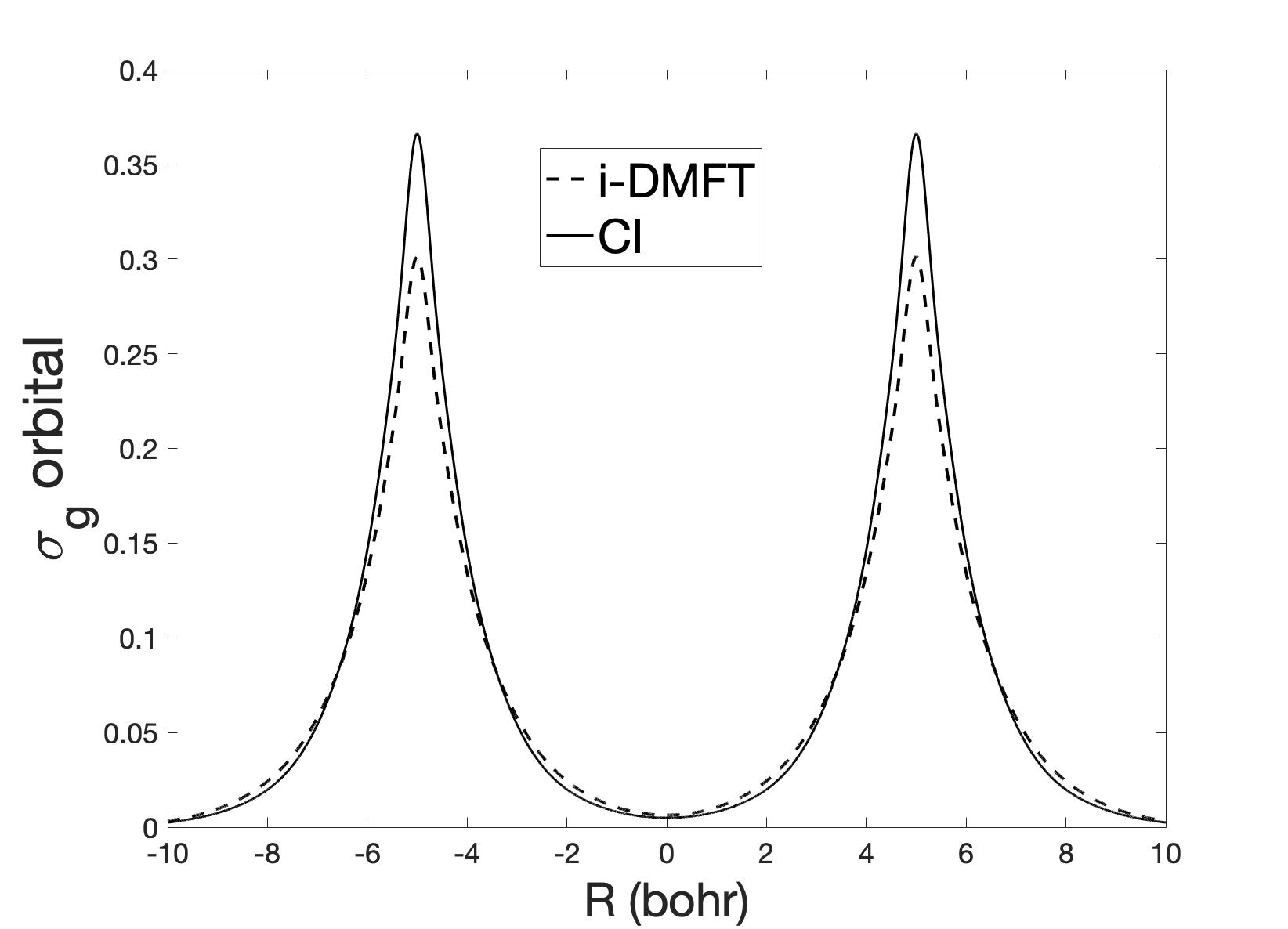}
 \end{center}
  \vspace{-0cm}
\caption{The $\sigma_g$ orbital from the $\it i$-DMFT calculation and the $\sigma_g$ NO from a CI calculation at $R$=10 bohr with the basis set cc-pVDZ. The $\it i$-DMFT parameters are $\kappa$=0.094681 and $b$=0.0286189 (in a.u.).}
\label{fig:orbsH2DMFTCI}
\end{figure}

DLS question our results for the dissociation of molecules using the example of the two-electron molecules H$_2$, HeH$^+$ and He$_2^{2+}$. For H$_2$ and He$_2^{2+}$ they do find, using exact NOs and ONs, the remarkable linear relationship \cite{WangPRA2021,WangPRA20212} of $E_{cum}$ vs. $S(\mathbf{n})$ which was the inspiration for our approach. However, they stress that the 1DM we obtain for H$_2$  differs from the exact/FCI 1RDM, judging from the deviation of the Frobenius norm $\sqrt{{\rm Tr}\textcolor{black}{(\gamma^{i}-\gamma)^2}}$ from zero, \textcolor{black}{where $\gamma^i$ is from $i$-DMFT and $\gamma$ form CI calculations}. This is especially true at the dissociation limit.

As discussed in (i), a difference of the 1DM of the $i$-DMFT calculation from the 1RDM of the exact wave function is in general to be expected.  
This is certainly the case for H$_2$, as already detailed in the  supplementary information (SI)  of  \cite{WangPRL2022}. It is useful to expand on this point, also in connection with the complaint of DLS about the nonuniversality of our functional, which manifests itself in the different values for $\kappa$ and $b$ for different molecules (even iso-electronic ones). \\

\begin{figure}[h]
 \begin{center}
  \includegraphics[width=.6\textwidth]{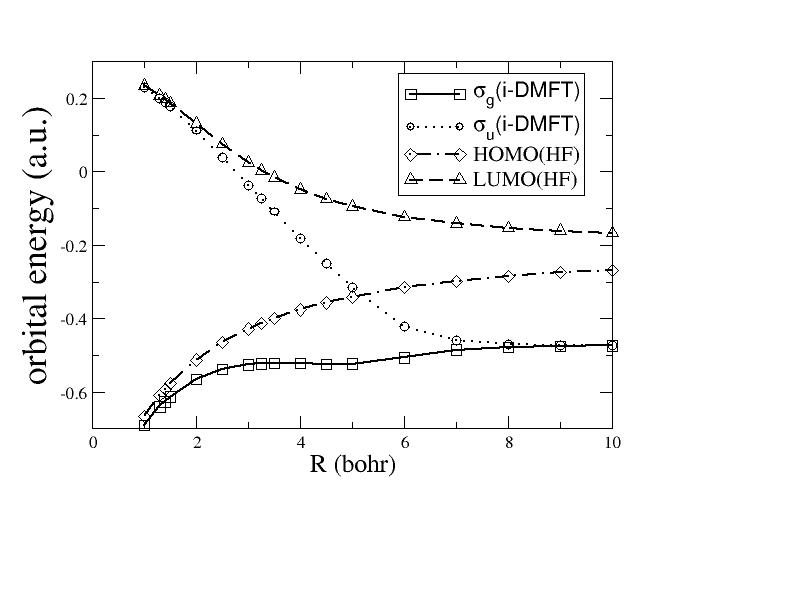}
 \end{center}
  \vspace{-1.8cm}
\caption{The orbital energies of H$_2$ with the entropic functional including the correction with the exchange energy as factor, Eq.\ \eqref{eq:ExS} with A=1/(4 $\rm ln$2). For comparison the HF orbital energies are also shown, see also Fig. 1 of SI in Ref.\ \cite{WangPRL2022}. }\label{fig:orbennewEc}
\end{figure}

It is well-known that the fundamental problem of the HF model is the presence of ionic terms in the wavefunction, which lead to too much on-site electron-electron repulsion in general and particularly at the dissociation limit. When substituting the HF 1RDM 
\begin{equation}\label{eq:gammaHF}
\gamma^{HF}(1,1')= \sum_{\sigma=\alpha,\beta} 1.0\ \sigma_g(\mathbf{r}_1)\sigma_g^*(\mathbf{r}'_1)\sigma(s_1)\sigma^*(s'_1)
\end{equation}
into the HF expression for the electron-electron energy, $Y$, the ionic terms show up when expanding the $\sigma_g$ orbital in the atomic orbitals (AOs), $\sigma_g=(\textcolor{black}{1/\sqrt 2})(1s_L+1s_R)$. DLS make the point that when the exact 1RDM is used,
\begin {eqnarray}\label{eq:gammaCI}
&\gamma^{CI}(1,1')=\sum_{\sigma=\alpha,\beta} 0.5\  \sigma_g(\mathbf{r}_1)\sigma_g^*(\mathbf{r}'_1)\sigma(s_1)\sigma^*(s'_1) \notag \\
 & \qquad +         \sum_{\sigma=\alpha,\beta} 0.5\  \sigma_u(\mathbf{r}_1)\sigma_u^*(\mathbf{r}'_1)\sigma(s_1)\sigma^*(s'_1)  \\
  & = \sum_{\sigma=\alpha,\beta} 0.5\ \left[1s_L(\mathbf{r})1s_L(\mathbf{r}')+1s_R(\mathbf{r})1s_R(\mathbf{r}') \right]\sigma(s_1)\sigma^*(s'_1) \notag
\end{eqnarray}  
the $Y$ of Eq.\ \eqref{eq:Y} with $\gamma^{CI}$ would again yield wrong on-site electron repulsion terms. They ascribe this to the electron-electron energy $E_{ee}$ of Eq.\ \eqref{eq:Eee} referring only to the regime of weak correlations.  But Eq.\ \eqref{eq:Eee} is remarkably successful for strong correlation. It is just the purpose of the entropic term in Eq.\ \eqref{eq:Eee} to correct for the correlation error also and in particular for strong (nondynamical) correlation.  The excellent total energies that are obtained with $i$-DMFT, along the complete dissociation coordinate, prove that the entropic term does have this effect.

Actually, the deviation of the Frobenius norm from zero has a different origin.  As has been discussed at some length in our paper  \cite{WangPRL2022} and the SI, there is an error in the orbitals and the electron density for H$_2$, notably at the dissociation limit. It is worthwhile to highlight this point here. We have stressed that our SCF equation, being very similar to the HF one, leads to the too diffuse orbitals and density that are characteristic of the HF model due to the too high on-site electron-electron repulsion (making the electronic potential part of the Fock operator too repulsive) \cite{BaerendsGritsenkoJPCA1997,GritsenkoSchipperBaerends1997}. That manifests itself in the too high orbital energy (not negative enough), see Table V and Fig.\ 1 of SI.  Such an orbital energy implies a too slow asymptotic decay. The diffuse nature of the HF $\sigma_g$ orbital, and the $i$-DMFT $\sigma_g$ (and $\sigma_u$)  orbitals compared to the $\sigma_g$ (and $\sigma_u$) NOs is evident from Fig.\ \ref{fig:orbsH2DMFTCI} of this work. The orbital energy at the dissociation limit should be $-0.5$ a.u.\ in order to have the correct shape and decay like the density of a H atom. 

The too diffuse orbitals and density show up in the Frobenius norm
\begin{eqnarray}\label{eq:Frobenius}
{\rm Tr} (\gamma^{i}-\gamma)^2 &=& \sum_{\mu} <\chi_{\mu}|(\gamma^i-\gamma)^2|\chi_{\mu}> \nonumber \\
&=&\sum_{\mu} <\chi_{\mu}|(\gamma^i)^2+\gamma^2-\gamma^i\gamma-\gamma\gamma^i|\chi_{\mu}>
\end{eqnarray} 
where $\{\chi_\mu\}$ is some complete basis.  
The terms $(\gamma^i)^2$ and $\gamma^2$ can be reduced to the sums over the squares of the occupation numbers by choosing for the basis either the NOs (for $\gamma^2$) or the eigenfunctions of $\gamma^i$ (for $(\gamma^i)^2$). As shown in, for example, Fig. 2 of our paper \cite{WangPRL2022} those two sets of  occupation numbers are very close, so these terms are not the source of the Frobenius norm discrepancy. However, the cross terms
$\gamma^i\gamma$ and $\gamma\gamma^i$ depend on the orbitals from two different calculations.
Choosing the set $\{\chi_{\mu}\}$ to be the NOs of the wave function, $\gamma \chi_{\mu}=n_{\mu} \chi_{\mu}$, 
and inserting the identity as a sum over eigenfunctions of $\gamma^i$, 
\begin{eqnarray}
\sum_{\mu,k} <\chi_{\mu}|\gamma^i | \psi_k><\psi_k|\gamma|\chi_{\mu}>=\sum_{\mu,k} n_k^i S_{\mu k}n_{\mu} S_{k \mu}
\end{eqnarray}
we see that the trace depends on the overlaps $S_{\mu k}$ of the two orbital sets from different calculations. Due to the different shapes of the $i$-DMFT orbitals and the exact NOs, the overlaps are not delta functions and the Frobenius norm will differ from zero.

\begin{figure}[h]
 \begin{center}
  \includegraphics[width=.5\textwidth]{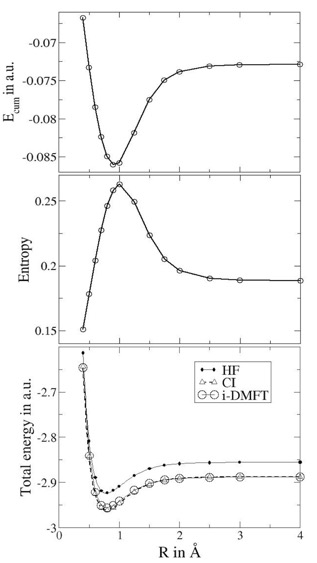}
 \end{center}
  \vspace{-0cm}
\caption{The upper and middle panels show the  cumulant energy and the entropy along the internuclear distance in HeH$^+$ from the analysis of the CI wave function. The lower panel shows the total energies as a function of internuclear distance from different methods. The basis set cc-pVDZ is used and for  
the $\it i$-DMFT calculation, $\kappa$=0.015 and $b$=0.03244 (in a.u.).}
%solid line in the lower panel is the linear regression with $E_{\rm cum}$= - 0.17228 $S$ - 0.040938.}
\label{fig:HeH+}
\end{figure}

\begin{figure}[h]
 \begin{center}
  \includegraphics[width=.6\textwidth]{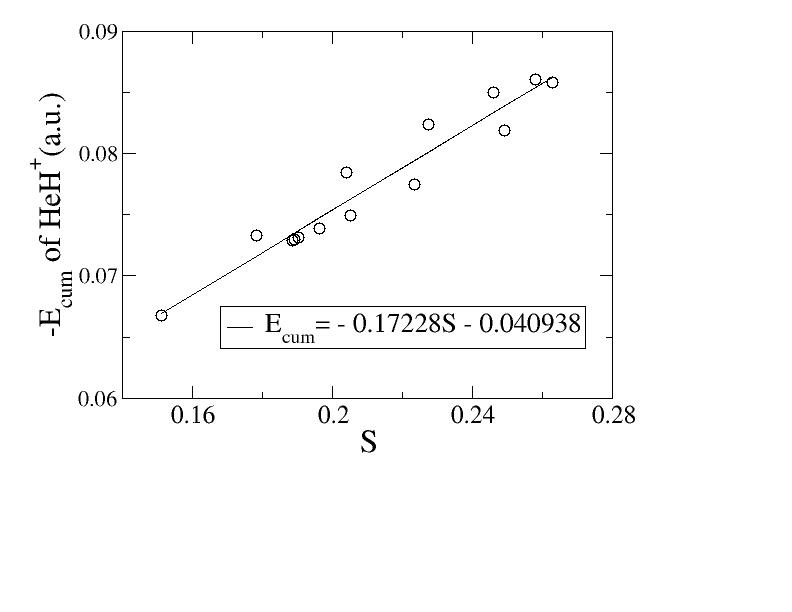}
 \end{center}
  \vspace{-2cm}
\caption{Linear regression to the points $E_{cum}$ vs $S$ for HeH$^+$ from CI calculations. The points below the line correspond to the long distance regime (beyond 1 $\AA$), the points above the line belong to the short distance regime (below 1 $\AA$).}
%solid line in the lower panel is the linear regression with $E_{\rm cum}$= - 0.17228 $S$ - 0.040938.}
\label{fig:S_Ecum}
\end{figure}

\begin{table}
\caption{\label{tab:densities}The entropy, cumulant energy, HF, CI energies in HeH$^+$ from the wave functions calculated with the basis set cc-pVDZ. The $\it i$-DMFT energies are obtained with $\kappa$=0.015 and $b$=0.03244 (in a.u.).}
\begin{ruledtabular}
\begin{tabular}{lcccccc}
   $R$ &    $S$ &   $E_{\rm cum}$ &HF & CI & $\it i$-DMFT \\ \hline
0.40 &  0.150975 & -0.066769 & -2.613107 & -2.643258 & -2.645547\\
0.50 &  0.178110 & -0.073258 & -2.809079 & -2.841901 & -2.841519\\
0.60 &  0.204076 & -0.078466 & -2.889631 & -2.924526 & -2.922072\\
0.70 &  0.227434 & -0.082393 & -2.918503 & -2.954909 & -2.950943\\
0.80 &  0.246080 & -0.084944 & -2.923532 & -2.960888 & -2.955972\\
0.90 &  0.258079 & -0.086032 & -2.918065 & -2.955797 & -2.950506\\
1.00 &  0.262753 & -0.085802 & -2.908716 & -2.946316 & -2.941156\\
1.25 &  0.249292 & -0.081863 & -2.885203 & -2.921216 & -2.917643\\
1.50 &  0.223434 & -0.077492 & -2.869870 & -2.904148 & -2.902311\\
1.75 &  0.205242 & -0.074961 & -2.862084 & -2.895344 & -2.894524\\
2.00 &  0.196277 & -0.073826 & -2.858529 & -2.891332 & -2.890969\\
2.50 &  0.190328 & -0.073105 & -2.856088 & -2.888605 & -2.888528\\
3.00 &  0.188979 & -0.072941 & -2.855461 & -2.887913 & -2.887896\\
4.00 &  0.188581 & -0.072894 & -2.855214 & -2.887648 & -2.887648\\   
\end{tabular}
\end{ruledtabular}
\end{table}
 
To remedy this particular HF error in the orbitals (for which dissociated H$_2$ is the worst case), an expedient modification of our functional is possible. Also the nonuniversality of our functional is an issue: the system dependence of the $\kappa$ parameter should eventually be replaced by functionals of the orbitals and occupation numbers (density matrix) in order to arrive at a universal functional. As an example of how to address these points we introduce orbital dependence into the entropic term through the exchange energy as a factor,
\begin{equation}\label{eq:ExS}
E_{cum}=AE_xS-b
\end{equation}
 where $E_x=-\frac{1}{2}\sum_{ij}n_i n_j<ij|ji>$. This leads to a new eigenvalue problem for the orbitals,
 \begin{eqnarray}\label{eq:Fchi}
\hat{F} \chi_i(1) = \epsilon_i \chi_i (1)
\end{eqnarray}
where the operator $\hat{F}$ is, 
\begin{eqnarray}\label{eq:F}
\hat{F} = \hat{h}+\sum_{j} n_j [\hat{J}_j- (1+A S)\hat{K}_j].
\end{eqnarray}
The occupation numbers are approximated with a Fermi-Dirac type of dependence on the orbital energies.

As shown in Fig.\ \ref{fig:orbennewEc} the orbital energies of the $\sigma_g$ and $\sigma_u$ orbitals now approach correctly the value of $-0.5$. This is because the electronic potential in the new Fock operator no longer includes the erroneous on-site repulsion of HF (when the reference electron is at site A, there is still 1/2 of the other electron at that site in the HF model). It  has electron correlation built in in the sense that when the reference electron is at site A, the other electron is fully at site B. This new Fock operator incorporates the potential of the full exchange-correlation hole, not just the exchange hole. It shares this property with the exact Kohn-Sham potential. The reference electron for which the orbital is determined by Eq.\ \eqref{eq:Fchi}, then sees when at site A  the nucleus A unscreened and the electron density acquires the correct shape of the H atom density at site A (and similarly at site B). This remedies the main source of the discrepancy signaled by the Frobenius norm differing from zero at long internuclear distance.  This development also points the way to reducing the nonuniversality of the $i$-DMFT functional. Of course the functional of Eq.\ \eqref{eq:ExS}, with a system dependent constant $A$, has deficiencies and is certainly not yet universal. It is just a first example of the further refinement of the $i$-DMFT functional with density and orbital dependent terms.
Ultimately one would strive for any remaining constants to be system independent, or, as is common practice in present day density functional development, to be determined by minimum deviations over appropriate benchmark sets of molecules.\\
\\  
DLS also signal a special problem for the HeH$^+$ molecule: they state that for HeH$^+$ there is not a single-valued dependence of $E_{cum}$ on $S$ and ``thus $i$-DMFT would fail to describe its chemical behavior''. In reality there is no problem with HeH$^+$. 
The upper and middle panels of Fig.\ \ref{fig:HeH+} clearly show an impressive scaling behaviour between $E_{\rm cum}$ and the entropy, $E_{\rm cum} \sim -\kappa S$. 
In Fig. \ref{fig:S_Ecum} we show the linear regression for $E_{cum}$ vs $S$ with the data points obtained from a CI calculation (see data in Table I), as in the middle panel of Fig.\ 1 of DLS. Note that the points below the straight line belong to the long distance regime (beyond 1 $\AA$), and the points above the straight line belong to the short distance regime (below 1 $\AA$). In both regimes, there is a close linear relation,  with a similar slope, between $E_{cum}$ and $S$ calculated from the CI wavefunction.  This is the interpretation of the middle panel of DLS' Fig.\ 1.

 It is to be noted that the chemical behavior of HeH$^+$ is different from the dissociations of covalent bonds in H$_2$, He$_2^{2+}$. At long internuclear distance this molecule dissociates to a proton (H$^+$) and a He atom (slightly polarized by the proton). Then we do not have large nondynamical correlation, like in the covalently bound molecules at long distance, but only the dynamical correlation of the He atom. When the H$^+$ approaches the He atom, some charge delocalization towards H$^+$ takes place and there is at shorter distance than ca.\ 1 $\AA$ a covalent component (or rather donor-acceptor component) to the bonding. In contrast to dissociating covalent bonds, like H$_2$ and He$_2^{2+}$, HeH$^+$ is pretty well described by the HF model along the entire dissociation coordinate. So both $E_{cum}$ and $S$ are relatively small. This explains why the axis scale for HeH$^+$ in the middle panel of Fig.\ 1 of DLS (both for $S$ and for $E_{cum}$) is almost an order of magnitude smaller than that in the left panel for the covalently bonded H$_2$ and He$_2^{2+}$. The differences in slope of the $E_{cum}$ vs $S$ lines in the two different distance and bonding \textcolor{black}{regimes shown} in Fig.\ \ref{fig:S_Ecum} 
are not large and a good dissociation curve is obtained with just one choice for $\kappa$ in Fig.\ \ref{fig:HeH+}, bottom panel.  
%The bottom panel of Fig.\ \ref{fig:HeH+} shows that the $E$ vs $R$ curve for HeH$^+$ has the same quality as in the other cases we have investigated.
Rather than failing for such a special system, with different bonding behavior in different distance regimes, the $i$-DMFT method can evidently cope with this situation. \\
\\ 
(iii) 
DLS state that in the dissociation limit of H$_2$ other theoretical forms would serve better than our entropic $S(\mathbf{n})$, with the weak condition that they be Schur-convex. There is no objection to other forms, of course, but the challenge is to prove that such a form works well over the entire range of the dissociation coordinate where the requirement is not just simple ``maximal mixedness'' of a limited set of orbitals. 
We reiterate that, since our orbitals are not NOs, the small occupation numbers are not expected to be equal to NO occupation numbers, and a scaling condition like $\partial \mathcal{F}/\partial n_i \sim -1/\sqrt{n_i}$ does not apply.

%\bibliography{../../../../../Artikelen/Libraries/biblioTCVU2016}
%\bibliography{Lagrange10}

\end{document}